\documentclass{ws-ijmpd}
\begin{document}

\markboth{Davood Momeni}
{Covariant Quantum-Mechanical Scattering  via Stueckelberg-Horwitz-Piron Theory}

%
\catchline{}{}{}{}{}
%

\title{COVARIANT QUNATUM-MECHANICAL SCATTERING VIA STUECKELBERG-HORWITZ-PIRON THEORY
  }

\author{DAVOOD MOMENI
}

\address{{Department of Physics, College of Science, Sultan Qaboos University,P.O. Box 36, ,AL-Khodh 123  Muscat, Oman}\footnote{Tomsk State Pedagogical University, TSPU, 634061 Tomsk, Russia } \\ \footnote{ Center for Space Research, North-West University, Mafikeng, South Africa} \\
davood@squ.edu.om}

\maketitle

\begin{history}
\received{Day Month Year}
\revised{Day Month Year}
\end{history}

\begin{abstract}
Based on the Stueckelberg-Horwitz-Piron theory of covariant quantum mechanics on curved spacetime, we solved wave equation for a charged covariant harmonic oscillator in the background of  charged static spherically symmetric black hole. Using Green’s functions , we found asymptotic form for the wave function in the lowest mode (s-mode) and in higher moments. It has been proven that for s-wave, in a definite range of solid angles, the differential cross section depends effectively to the magnetic and electric charges of the black hole.
\end{abstract}

\keywords{Covariant quantum theory on curved space, Stueckelberg-Horwitz-Piron theory.}

\ccode{PACS numbers:03.30.+p, 03.65.-w, 04.20 Cr, 04.60 Ds, 04.90.+e}


\section{Introduction}

General relativity (GR) considered as the best classical  gauge theory to explore gravitational interaction locally and globally  \cite{Moshe}. Several tests provided a trustable framework for GR. Because quantum mechanics (QM) also proved to be the best non relativistic framework to study tiny scales systems, it is naturally arised a question that whether QM can be written in a covariant form to include all qunatum effects on curved space time manifolds or not?. Since GR has locally Lorentz invariance, as a result to build covariant QM, we need first make QM relativistic and it is adequate to make it in canonical formalism. It was Stueckelberg who bulit 
a relativitic quantum mechanics in canonical formalism \cite{Stueckelberg}. Since relativistic wave equations have always many body interpretations, it was adequate to generalize the Stueckelberg's works to such many body cases. A generalization of the Stueckelberg's relativistic canonical formalism 
to  many body systems investigated by Horwitz and Piron\cite{SHP}. This theory  which we will refer to as SHP, recently   revisited  by Horwitz  and generalized it to the general Lorentzian curved manifold  \cite{arXiv:1810.09248}. In a recent paper I investigated exact solutions for SHP theory both in classical and quantum domains for a covariant simple harmonic oscilator (CSHO) in the vicinity of a black hole \cite{Momeni:2019knc}. In continuation of my recent study and because SHP theory plays an important role to build covariant QM in curved space times, in this work I studied exact solutions for a charged CSHO in the background of magnetically Charged  black hole. I allow that the CSHO have also a background depedent interaction with an external gauge field. The first application of the results in this letter will be a starting point to investigate covariant 
Aharonov
and Bohm effect on curved backgrounds. It is well knowm that using the Schr\"odinger equation, the problem of the
scattering of an electron in an external static magnetic field, in flat space showed an independent 
depth of penetration of the electrons into  region of non zero magnetic force
lines. This effect discovered in 1959 by Aharonov
and Bohm  \cite{AB1} and named as Aharonov
-Bohm (AB) effect. An direct interpretation of AB is that 
the external $U(1)$ electromagnetic field
interacts with  the  charged particles and penetrateto to the  region in which the
field is localized and according to the QM , cannot be reached by the particles (see,
the reviews of Refs. \cite{Peshkin} and \cite{Afanas'ev}). AB effect in curved space proposed in  past  Ref. \cite{Sitenko:1997uc} but in that interesting study the wave equation for QM didn't preserve general covariance. Consequently still it is very important to 
investigate the effect of curvature  space on the
Aharonov-Bohm As an starting point, in this letter I considered covariant QM wave equation proposed in the SHP theory  and 
I consider the situation in which
there is an external  magnetic force as well as an additional 
 static spherically  symmetric gravitational field. I will solve the wave equation and will investigate the possible lower and higher modes scattering of this particle from the horizon of  black hole. Note that speed of light  $c=1$ in our units convention through whole analysis in this letter.

\section{Quantum mechanics via Schr\"{o}dinger-Stueckelberg -Horwitz-Piron wave equation on Magnetically Charged Black Holes}
According to the pionerring work of Horwitz, the covariant 
quantum mechnaical wave equation for  particle  on a curved general relativity background well formulated in \cite{ arXiv:1810.09248} and  as the below Schr\"{o}dinger-Stueckelberg -Horwitz-Piron wave equation:
\begin{eqnarray}
&&i\hbar\frac{\partial}{\partial\tau}\Psi_{\tau}(x^\mu)=\hat{H}\Psi_{\tau}(x^\mu)\label{waveeq}
\end{eqnarray}
In the analogous to the QM, 
$\hat{H}$ is quantum mechanical operator. In SHP theory there are two types of the time coordinates. First is the coordinate time $t$ as a component of the four vector coordinates $x^{mu}$. This time $t$ is a dynamical variable on the manifold $\mathcal{M}$. To understand the QM , we need another time,the chronological or historical time $\tau$. We use this time to record the history of an event happened in the manifold. The  chronological time $\tau$ corresponds to the time revolution of the physical Hamiltoninan in the manifold.
The wave equation presented in Eq. (\ref{waveeq}) defines 
a Hilbert space with scalar product as follows:
\begin{eqnarray}
&&(\psi,\chi)=\int d^4x\sqrt{-g} \Psi^{*}_{\tau}(x^\mu)\chi_{\tau}(x^\mu)\label{inner}
\end{eqnarray}
In the above definition, 
the volume element is written as $d^4x\sqrt{-g}$ and $^{*}$ denotes complex conjugatre. The appropriate form for a quantum mechanical Hamiltonian $\hat{H}$ by  following convention of indexes given in ref.  \cite{arXiv:1810.09248} defined   as:
\begin{eqnarray}
&&\hat{H}=\frac{1}{2M\sqrt{-g}}(p_{\mu}-a_{\mu})g^{\mu\nu}(p_{\nu}-a_{\nu})+V(x)\label{H1}.
\end{eqnarray}
Very recently we have investigated exact mode decomposition solutions for wave equation (\ref{waveeq}) in \cite{Momeni:2019knc}. The aim in this letter is to use this covariant wave equation on curved spacetime built on a spherically symmetric-stationary  magnetic-electric charged black hole  metric in the Schwarzschild coordinates as follows\cite{Weinberg:1994eb},
\begin{eqnarray}
&&ds^2=-B(r)dt^2+B^{-1}(r)dr^2+r^2d\Omega^2
\label{g}.
\end{eqnarray}
here $B(r)=1-\frac{2MG}{r}+\frac{4\pi G}{r^2}\sqrt{Q_E^2+Q_M^2}$. In the above metric function, $Q_E,Q_M$ are electric and magnetic $U(1)$ charges. The corresponding fields due to the metric (\ref{g}) are as follow:
\begin{eqnarray}
&&E_r=\frac{Q_E}{r^2}.\label{E}\\ &&B_r=\frac{Q_M}{r^2}\label{B}.
\label{g}.
\end{eqnarray}
Note that the geometrical structure of the metric (\ref{g}) can be described by defining of an extremal mass parameter as follows: 
\begin{eqnarray}
&&M_{ext}=\sqrt{4\pi(Q_E^2+Q_M^2)}.M_{pl}
\label{M}.
\end{eqnarray}
here $M_{pl}=G^{-1/2}$.
There are three different cases for geometry of metric :
\begin{itemize}
	\item if $M>M_{ext}$, there is a pair of real zeros for algebraic equation $B(r)=0$, given by $r_{\pm}$(singularities). The region outside $r>r_{+}$ , is a region out of the horizon.
	\item if $M=M_{ext}$, the radius $r_{+}=r_{-}$ coincides, we end up by an exttremal Reissner-Nordstrom  blackhole.
	\item If $M<M_{ext}$, the spacetime has a naked singularity, it implies a no horizon solution. We obey the cosmic censorship consequaently we aviod from the naked singularity.  
\end{itemize}
Though this study we will condsider  $M>M_{ext}$ as our possible black hole background.\par
It is adequate to mention  here about the role of Dirac-string when the magnetic field eq. (\ref{B}) arises from a monopole with  charge $Q_M$ gives rise to a radial $\vec{B}=B_r \hat{r}$. The magnetic field (\ref{B}) yields to a vector potential $\vec{A}^{Dirac}\equiv \vec{A}$. Using the selonoid condition, $\nabla\cdot\vec{B}=0$, we have $\vec{B}=\nabla\times \vec{A}$, in the coordinates $x^{\mu}=(r,\theta,\varphi)$ we can opt the vector potential as follows:
\begin{eqnarray}
&&\vec{A}=(0,0,Q_M(1-\cos\theta))
\label{A}.
\end{eqnarray}
The above vector potenatial eq. (\ref{A}), has a singularity for $z(=r\cos\theta)<0$, called Dirac singularity. A suitable gauge transformation can move (but not remove)  this singularity.
\par
In flat space, the  AB effect makes sense to this string unobservable. We can write $Q_M$ using Dirac qunatization as
\begin{eqnarray}
&&Q_M=\frac{Q_E}{e}\in \mathcal{Z} \ \ \mbox{or}\ \  \mathcal{Z}+\frac{1}{2}.
\label{qm}.
\end{eqnarray}
Note that still we have a singularity at $r=0$. This  can be explained via a spontaneously broken gauge have non singular classical solutions,
\begin{eqnarray}
&&SU(2)\to U(1)\times \phi_{Higgs}.
\end{eqnarray}
where the vacuum expectation value of the Higgs field $<\phi_{Higgs}>\neq 0$. In our study on curved spacetime which is based on the  charged black hole solution (\ref{g}), the aim is to split the Hamiltonian to $\hat H=\hat H_0+\Delta \hat H$, here $\Delta \hat H$ is the perturbation term. I will compute the scattering cross section as well as exact solutions for unperturbated part $\hat H_0$. We use an ansatz $a_{\mu}=(a(r),\vec{A}(\theta))$, where $a(r)$ is the scalar electric potential and $\vec{A}$ is the Dirac vector potentail in eq. (\ref{A}). Following our former study in \cite{Momeni:2019knc}, we consider the potential $V(r)$ as a covariant harmonic oscilator. In the background metric (\ref{g}),the decomposition of equation  (\ref{H1}) with the below form is possible:

\begin{eqnarray}
&&\Psi_{\tau}(x^\mu)=\exp\Big[-\frac{iE\tau}{\hbar}-i\omega t
\Big]\Phi(r,\theta,\varphi)\label{wave solution}
\end{eqnarray}
In the Lorentz's gauge, 
\begin{eqnarray}
&&\partial_{\mu}\Big(g^{\mu\nu}a_{\nu}
\Big)=0.
\end{eqnarray}
This gauge fixing condition gives us $a(r)=\frac{Q_E}{r}$ as a coulomb's scalar potential in the theory.
We will end up by the folllowing partial differential equation, eq. (\ref{wave solution}),

	\begin{eqnarray}
	&&\frac{1}{r^2}\frac{\partial}{\partial r}\Big(r^2\frac{\partial \Phi(r,\theta,\varphi)}{\partial r}\Big)+\frac{1}{r^2\sin\theta}\frac{\partial}{\partial \theta}\Big(\sin\theta\frac{\partial \Phi(r,\theta,\varphi)}{\partial \theta}\Big)+\frac{1}{r^2\sin^2\theta}\frac{\partial^2\Phi(r,\theta,\varphi)}{\partial \varphi^2}\nonumber \\&&+\frac{1}{r^2}\frac{\partial}{\partial r}\Big(r^2\delta B\frac{\partial \Phi(r,\theta,\varphi)}{\partial r}\Big)-\Big[\frac{2\omega a(r)}{B(r)}+\frac{2\hbar \mu A(\theta)}{r^2\sin^2\theta}-E
	+V(r)\Big]\Phi(r,\theta,\varphi)=0.\label{Phi}
	\end{eqnarray}

where we rewrite the metric function in the weak regime as follows,
\begin{eqnarray}
&&B(r)=1+\delta B(r),\ \ |\delta B(r)|\ll1.
\end{eqnarray}
in our case, $\delta B(r)=-\frac{2GM}{r}+\frac{GM_{ext}}{\sqrt{4\pi}M_{pl}}\frac{1}{r^2}$. We can rewrite eq. (\ref{Phi}), in the following operator form in the representation theory,
\begin{eqnarray}
&&\hat{H}_0\Phi+\epsilon(\Delta \hat H)\Phi=0\label{totalH}.
\end{eqnarray}
where in our perturbative representation , we can write the unperturbated Hamiltoninan in terms of the quantum operator $\vec{\hat L}$.
\begin{eqnarray}
&&\hat{H}_0=-\frac{\vec{\hat L}^2}{\hbar^2}
\end{eqnarray}
and 

	\begin{eqnarray}
	&&\epsilon\Delta \hat H=\frac{1}{r^2}\frac{\partial}{\partial r}\Big(r^2\delta B\frac{\partial }{\partial r}\Big)-\Big[\frac{2\omega a(r)}{B(r)}+\frac{2\hbar \mu A(\theta)}{r^2\sin^2\theta}-E
	+V(r)\Big]
	\end{eqnarray}

for CSHO , the scalar potential is given as  $V(r)=\frac{1}{2}m\Omega^2 r^2$ with mass $m$, the frequency $\Omega$ nad radial coordinate $r$. We solve the unperturbated wave equation, the below differential equation: 
\begin{eqnarray}
&&\hat{H}_0\Phi^{0}=0
\end{eqnarray}
An exact solution is expressed in terms of the radial and spherical harmonics,
\begin{eqnarray}
&&\Phi^{0}=\sum_{l,\mu}c_{l\mu}Y_{l\mu}(\theta,\varphi) r^{-(l+1)}\label{phi0}
\end{eqnarray}
Using the iteration technique, we need to substitue $\Phi^{0}$ in $\epsilon(\Delta \hat H)\Phi$, and taking the first approximation, we obtain:
\begin{eqnarray}
&&\Phi^{1}=-\epsilon H_0^{-1}(\Delta \hat H)\Phi^0\label{Phi1}. 
\end{eqnarray}
Now to solve the total Hamiltonian wave eqaution, i.e,Eq. (\ref{totalH}) we use iteration method. We take the unperturbated solution (\ref{phi0})as zeroth order approximation and by inserting it to the full perturbated system Eq. (\ref{totalH}), 
on the first level of the perturbation theory, we obtain:

	\begin{eqnarray}
	&&\Phi(\vec{r})\approx \Phi^0(\vec{r})+\epsilon\int_{\Omega} G(\vec{r},\vec{r}')H_0^{-1}(\Delta \hat H(\vec{r}'))\Phi^0(\vec{r}')d^3r'\label{s}
	\end{eqnarray}

A general exact solution for (\ref{Phi1})

	\begin{eqnarray}
	&&\Phi^{1}=\epsilon\int_{\Omega} G(\vec{r},\vec{r}')\Delta \hat H(\vec{r}')\Phi^0(\vec{r}')d^3r'\label{phitot}
	\end{eqnarray}

here $\vec{r}'\equiv (r',\theta',\varphi'),d^3r'=r'^2\sin\theta'dr'd\theta'd\varphi'$. The appropriate  Dirichlet Green’s function for
the region 
$\Omega$ bounded by $r_{+}\leq r<\infty$ obtained 
in terms of the harmonic functions of the Laplace operator $\Delta=\hat{H}_{0}^{-1}$ as follows:

	\begin{eqnarray}
	&&G(\vec{r},\vec{r}')=\sum_{l=0}^{\infty}\sum_{\mu=-l}^{+l}\frac{4\pi}{2l+1}Y_{l\mu}(\theta,\varphi)Y^{*}_{l\mu}(\theta',\varphi')\Big[\frac{r_{<}^l}{r_{>}^{l+1}}-\frac{r_{+}^{2l+1}}{(r_{<}r_{>})^{l+1}}
	\Big]
	\label{green}
	\end{eqnarray}

where $r_{<,>}$ is the smaller (larger) of $r$ and $r'$.
The first order solution for $\phi^{1}$ is obtained by 
\begin{eqnarray}
&&\Phi^1(\vec{r})=\epsilon
\int_{\Omega} G(\vec{r},\vec{r}')\Delta \hat H(\vec{r}')\Phi^0(\vec{r}')d^3r'
\label{phi1}\end{eqnarray}
The next task is to calculate closed form for (\ref{phitot}) and specially study its lower order term , when $l=0$, called s-wave. Comparing $\Phi^{1}$ with the zeroth order approximated solution, provides a way to compute the amplitude between refrected wave to the incident wave.
\section{s-wave sacatering cross section}
Note that if we focus on s-wave, when $\Phi^0(\vec{r}')=\frac{a}{r'}$ where $a$ can be computed via the inner product defined in Eq.(\ref{inner}) with metric (\ref{g}), then $\epsilon\Delta \hat H(\vec{r}')\Phi^0(\vec{r}')=s_1(r')+s_2(r')s_3(\theta')
$ here the radial and azimutal functions represented as folllow:

	\begin{eqnarray}
	&&s_1(r)=-\frac{a\partial_r(\delta B(r))}{r^2}-\frac{a}{r}\Big[\frac{2\omega Q_E}{rB(r)}-E
	+\frac{1}{2}m\Omega^2 r^2\Big]\\&&
	s_2(r)=-\frac{2Q_Ma\hbar \mu }{r^3}\\&&
	s_3(\theta)=(1+cos\theta)^{-1}
	\end{eqnarray}

We can rewrite the source term, i.e., $\epsilon\Delta \hat H(\vec{r}')$ in terms of the spherical  harmonic functions according to the completness condition:

	\begin{eqnarray}
	&&\epsilon\Delta \hat H(\vec{r}')\Phi^0(\vec{r}')=\sqrt{4\pi}s_1(r')Y_{00}(\theta',\varphi')+s_2(r')\sum_{l'=0}^{\infty}\sum_{m'=-l'}^{+l'}a_{l'm'}Y_{l'm'}(\theta',\varphi')
	\end{eqnarray}

Note that since 

	\begin{eqnarray}&&
	a_{l'm'}=\int d\Omega's_3(\theta')Y^{*}_{l'm'}(\theta',\varphi')=\sqrt{\frac{2l'+1}{4\pi}}\sqrt{\frac{(l'-m')!}{(l'+m')!}}\delta_{m',0}\int_{-1}^{1} \frac{dx p_{l'}(x)}{1+x} 
	\end{eqnarray}

consequently the source term simplies to the follows:

	\begin{eqnarray}
	&&\epsilon\Delta \hat H(\vec{r}')\Phi^0(\vec{r}')=\sqrt{4\pi}s_1(r')Y_{00}(\theta',\varphi')+s_2(r')\sum_{l'=0}^{\infty}\sqrt{\frac{2l'+1}{4\pi}}p_{l'}(\cos\theta')c_{l'}\label{rho}
	\end{eqnarray}

here we define $c_{l'}\equiv \int_{-1}^{1} \frac{ p_{l'}(x)dx}{1+x}$, consequently plugging (\ref{rho}) in (\ref{phi1}) we obtain:

	\begin{eqnarray}
	&&\Phi^{1}=4\pi\int_{r_{+}}^{\infty}r'^2dr's_1(r')\Big[\frac{1}{r_{>}}-r_{+}	\Big]+\sum_{l=0}^{\infty}c_lp_l(\cos\theta) \sqrt{\frac{4\pi}{2l+1}}\int_{r_{+}}^{\infty}r'^2dr's_2(r')
	\Big[\frac{r_{<}^l}{r_{>}^{l+1}}-\frac{r_{+}^{2l+1}}{(r_{<}r_{>})^{l+1}}
	\Big]
	\end{eqnarray}

To calculate the integral we have to split it as follows:
\begin{eqnarray}&&
\int_{r_{+}}^{\infty}[...]dr'=\int_{r_{+}}^{r}[...]dr'+\int_{r}^{\infty}[...]dr'
\end{eqnarray}
using the above splitting we have:

	\begin{eqnarray}
	&&
	\int_{r_{+}}^{\infty}r'^2dr's_1(r')\Big[\frac{1}{r_{>}}-r_{+}	\Big]=\Big[\frac{1}{r}-r_{+}	\Big]\int_{r_{+}}^{r}r'^2dr's_1(r')+\int_{r_{+}}^{\infty}r'^2dr's_1(r')\Big[\frac{1}{r'}-r_{+}	\Big]\\&&
	\int_{r_{+}}^{\infty}r'^2dr's_2(r')
	\Big[\frac{r_{<}^l}{r_{>}^{l+1}}-\frac{r_{+}^{2l+1}}{(r_{<}r_{>})^{l+1}}
	\Big]=\int_{r_{+}}^{r}r'^2dr's_2(r')
	\Big[\frac{r'^l}{r^{l+1}}-\frac{r_{+}^{2l+1}}{(rr')^{l+1}}
	\Big]\\&&\nonumber+\int_{r}^{\infty}r'^2dr's_2(r')
	\Big[\frac{r^l}{r'^{l+1}}-\frac{r_{+}^{2l+1}}{(rr')^{l+1}}
	\Big]
	\end{eqnarray}

Finally the first order solution is obtained  in the closed form as follows:

	\begin{eqnarray}
	&&\Phi(\vec r)\approx\frac{a}{r}+4\pi\Big(\alpha^1_0(r_{+},r)+\beta^1_0(r_{+},r)\Big)
	+\sum_{l=0}^{\infty}c_lp_l(\cos\theta) \sqrt{\frac{4\pi}{2l+1}}
	\Big(\alpha^2_l(r_{+},r)+\beta^2_l(r_{+},r)\Big)
	\end{eqnarray}

here we define a set of auxiliary functions:

	\begin{eqnarray}
	&&
	\alpha^{1,2}_l(r_{+},r)=\int_{r_{+}}^{r}r'^2dr's_{1,2}(r')
	\Big[\frac{r'^l}{r^{l+1}}-\frac{r_{+}^{2l+1}}{(rr')^{l+1}}
	\Big]
	\\&&
	\beta^{1,2}_l(r_{+},r)=\int_{r}^{\infty}r'^2dr's_{1,2}(r')
	\Big[\frac{r^l}{r'^{l+1}}-\frac{r_{+}^{2l+1}}{(rr')^{l+1}}
	\Big]
	\end{eqnarray}

here $l=0...\infty$.
Note that in the scattering regime, when the fields are considered only at the asymptotic limit $r\to\infty$, 

	\begin{eqnarray}
	&&
	\alpha^{1,2}_l(r_{+},\infty)\approx\frac{1}{r^{l+1}}\int_{r_{+}}^{\infty}r'^2dr's_{1,2}(r')=\frac{\gamma^{1,2}_l(\infty)}{r^{l+1}}
	\\&&
	\beta^{1,2}_l(r_{+},r)\approx 0.
	\end{eqnarray}

and finally we have:

	\begin{eqnarray}
	&&\Phi(\vec{r})\approx \frac{a}{r}+\frac{4\pi\gamma^{1}_0(\infty)}{r}
	+\sum_{l=1}^{\infty}c_lp_l(\cos\theta) \sqrt{\frac{4\pi}{2l+1}}
	\frac{\gamma^{2}_l(\infty)}{r^{l+1}}
	\end{eqnarray}

The cross section for the scattered waves are

\begin{eqnarray}
&&\sigma(\theta,\varphi)\sin\theta d\theta d\varphi =|f(\theta,\varphi)|^2\sin\theta d\theta d\varphi.
\end{eqnarray}

here we have 
\begin{eqnarray}
&&
f(\theta,\varphi)=\sqrt{\frac{\pi }{3}}\frac{2  c_1
	\gamma^{2}_1(\infty)}{a^2} p_1(\cos\theta)
\end{eqnarray}

here 
\begin{eqnarray}
&&
\gamma^{2}_1(\infty)= 2Q_Ma\hbar \mu\ln \frac{r_{+}}{\Lambda}\\&& c_1=2\tanh^{-1}(\Lambda^{-1})
\end{eqnarray} 
and $\Lambda$ is an ultraviolet cutoff parameter. The total cross section for the s-wave scatering is given as follows:

\begin{eqnarray}
&&\sigma_{tot}=\frac{256 \pi ^2 \mu ^2 \hbar ^2
	\text{Arctanh}\left(\frac{1}{
		\Lambda 
	}\right)^2 Q_M^2 \log
	^2\left(\frac{r_{+}}{\Lambda
	}\right)}{9 a^2}.
\end{eqnarray}
here $r_{+}=1 + \sqrt{1 -\frac{2M_{ext}\sqrt{\pi}}{G M^2 M_{pl}}}$. If we find the normalization factor $a$ using the  integral (\ref{inner}) and by plugging the horizon radius $r_{+}$ finally we have:

	\begin{eqnarray}
	&&\sigma_{tot}=\frac{1024}{9} \pi ^3 \mu ^2 \hbar
	^2 Q_M^2
	\text{Arctanh}\left(\frac{1}{\Lambda
	}\right)^2 
	\left(\Lambda -1-\sqrt{1-\frac{2 \sqrt{\pi
			} M_{ext}}{G M^2
			M_{pl}}}\right)
	\ln
	^2\left(\frac{\sqrt{1-\frac{2 \sqrt{\pi
				} M_{ext}}{G M^2
				M_{pl}}}+1}{\Lambda }\right)
	\end{eqnarray}

In limit $\Lambda\to\infty$ it reduces to $\sigma_{tot}\approx\frac{1024 \pi ^3 \mu ^2 \hbar ^2
	Q_M^2}{9  }\frac{\log ^2\left(\Lambda
	\right)}{\Lambda}\approx 0$
A possible explanation may be the because black hole is magnetic as well, that might destroy the spherical symmetry and the s-wave cross section could vanish\cite{Horwitz2}. 
\section{General solution for $\Phi$ for higher orders moments}
In the previous  we focused on the lowest mode, when $\Phi^{0}(\vec r)=a r^{-1}$. In this section we wanna find a more general solution by inserting the  (\ref{phi0}). The aim is to find $\Phi^1(\vec r)$ by inserting $\Phi^0(\vec r)$ as the zeroth order solution. Firstly it is more siutable to write the  $\Phi^{0}(\vec r)$  and Green's function (\ref{green})in the following form:
\begin{eqnarray}
&&\Phi^{0}(\vec r)=\sum_{l=0}^{\infty}\sum_{\mu=-l}^{+l}f_{l\mu}(r)Y_{l\mu}(\theta,\varphi)\label{phi02}\\
&&G(\vec{r},\vec{r}')=\sum_{l=0}^{\infty}\sum_{\mu=-l}^{+l}g_{l\mu}(r,r')Y_{l\mu}(\theta,\varphi)Y^{*}_{l\mu}(\theta',\varphi')
\label{green2}
\end{eqnarray}
For our next purposes we mention here that 
\begin{eqnarray}
&&\partial_r\Phi^{0}(\vec r)=-\sum_{l=0}^{\infty}\sum_{\mu=-l}^{+l}(l+1)\frac{f_{l\mu}(r)}{r}Y_{l\mu}(\theta,\varphi)
\end{eqnarray}
Using the above expressions we can find the following source term:

	\begin{eqnarray}
	&&\epsilon\Delta \hat H(\vec{r}')\Phi^0(\vec{r}')=\sum_{l=0}^{\infty}\sum_{\mu=-l}^{+l}
	\Big(	E - \frac{1}{2} m \Omega^2 r^2 - \frac{2 Q_E \omega}{r (1 + \delta B(r))}-\frac{1}{r^2}  (l \delta B(r) - r \delta B(r)')\Big)\\&&\nonumber\times f_{l\mu}(r)Y_{l\mu}(\theta,\varphi)\\&&\nonumber+\sum_{l=0}^{\infty}\sum_{\mu=-l}^{+l}f_{l\mu}(r')Y_{l\mu}(\theta',\varphi') s_2(r')s_3(\theta')
	\end{eqnarray}

Note that $s_i$ are same functions as we defined in previous section. Note that still we have

\begin{eqnarray}
&&
s_3(\theta')=\sum_{l=0}^{\infty}\sum_{\mu=-l}^{+l}c_{l\mu}Y_{l\mu}(\theta,\varphi)\delta{\mu.0}
\end{eqnarray}

By plugging them in the $\Phi^1(\vec r)$, Eq. (\ref{phi1}) we obtain:

	\begin{eqnarray}
	&&\Phi^1(\vec{r})=\sum_{l=0}^{\infty}\sum_{\mu=-l}^{+l}\sum_{l'=0}^{\infty}\sum_{\mu'=-l'}^{+l'}\gamma_{ll'\mu\mu'}(r)Y_{l\mu}(\theta,\varphi)\int d\Omega'Y_{l'\mu'}(\theta',\varphi') Y^{*}_{l\mu}(\theta',\varphi')
	\\&&+\sum_{l,l',l''=0}^{\infty}\sum_{m,m',m''}c_{l''\mu''}\delta_{\mu'',0}\zeta_{ll'l''\mu\mu'\mu''}(r)Y_{l\mu}(\theta,\varphi)\int d\Omega'Y_{l'\mu'}(\theta',\varphi') Y^{*}_{l\mu}(\theta',\varphi')Y_{l''\mu''}(\theta',\varphi').
	\label{phi12}\end{eqnarray}

here

	\begin{eqnarray}
	&&\gamma_{ll'\mu\mu'}(r)=\int_{r_{+}}^{\infty} dr'^2 r'^2g_{l'\mu'}(r_{<},r_{>})f_{l\mu}(r')\Big(	E - \frac{1}{2} m \Omega^2 r'^2 - \frac{2 Q_E \omega}{r' (1 + \delta B(r'))}-\frac{1}{r'^2}  (l \delta B(r') - r'\partial_{r'} \delta B(r'))\Big)\\&&
	\zeta_{ll'l''\mu\mu'\mu''}(r)=\int_{r_{+}}^{\infty} dr'^2 r'^2g_{l'\mu'}(r_{<},r_{>})f_{l\mu}(r')s_2(r')\delta_{\mu'',0}
	\end{eqnarray}

remembering these identities for harmonic functions\cite{Arfken},

	\begin{eqnarray}
	&&\int d\Omega'Y_{l'\mu'}(\theta',\varphi') Y^{*}_{l\mu}(\theta',\varphi')=\delta_{ll'}\delta_{\mu\mu'}\\
	&&
	\int d\Omega' Y^{*}_{l\mu}(\theta',\varphi')Y_{l'\mu'}(\theta',\varphi')Y_{l''\mu''}(\theta',\varphi')=\sqrt{\frac{(2l'+1)(2l''+1)}{4\pi(2L+1)}}C(l,l',l''|0,0,0)C(l,l',l''|\mu,\mu',\mu'').
	\end{eqnarray}

Here $L=l+l'+l''$ and $C(l,l',l''|\mu,\mu',\mu'')$ is Clebsch-Goran coefficents. Such solution can be used to make a better approximation to the cross section obtained in previous. 

\section{Summary }
SHP theory provides a covariant quantum mechanical wave equation to study  mechanics on curved space times. A direct application can be a realization of the role   QM in the magnetic fields. In this letter I applied SHP theory in a general magnetic-electric charged background for a covariant harmonic oscilator as our quantum mechanical toy model. I solved the wave equation in lower and higher modes using perturbation (iteration) technique. I showed that in the s-mode,  total scatering cross section can be computed analytically by introducing a suitable ultraviolet cutoff parameter $\Lambda$. This UV cutoff can be imagined as the UV analogous to an infrared (near horizon) cutoff $\epsilon$. A remarkable observation was the total cross section vanishes at large values of cutoff parameter. We can interpret it as an isolation of the magnetic monopole by the dual electric charge in this charged black hole background. Furthermore I derived exact higher mode solution. In a forthcoming paper I will study Aharonov-Bohm effect using these covariance quantum mechanical wave solutions.

\section{Acknowledgment} 

I thank Prof. Lawrence P. Horwitz for carefully reading my first draft, very useful comments, corrections and discussions. 


\end{document}